\documentclass{article}

\usepackage[preprint, nonatbib]{neurips_2021}

\usepackage[utf8x]{inputenc} 
\usepackage[T1]{fontenc}    
\usepackage{url}            
\usepackage{booktabs}       
\usepackage{amsfonts}       
\usepackage{nicefrac}       
\usepackage{microtype}      
\usepackage{xcolor}         

\usepackage{graphicx}

\usepackage{subcaption}
\usepackage{multirow}
\usepackage{multicol}
\usepackage{amsmath,comment,listings}
\usepackage[ruled,vlined]{algorithm2e} 
\usepackage{algorithmic}

\title{Distributed SLIDE: Enabling Training Large Neural Networks on Low Bandwidth and Simple CPU-Clusters via Model Parallelism and Sparsity}

\author{%
  Minghao Yan \\
  Rice University\\
  \texttt{my29@rice.edu} \\
   \And
  Nicholas Meisburger \\
  Rice University and ThirdAI Corp.\\
  \texttt{ncm5@rice.edu} \\
   \AND
  Tharun Medini \\
  ThirdAI Corp. \\
  \texttt{tharun@thirdai.com} \\
   \And
  Anshumali Shrivastava \\
  Rice University and ThirdAI Corp.\\
  \texttt{anshumali@rice.edu} \\
}

\begin{document}

\maketitle

\begin{abstract}
More than 70\% of cloud computing is paid for but sits idle. A large fraction of these idle compute are cheap CPUs with few cores that are not utilized during the less busy hours. This paper aims to enable those CPU cycles to train heavyweight AI models. Our goal is against mainstream frameworks, which focus on leveraging expensive specialized ultra-high bandwidth interconnect to address the communication bottleneck in distributed neural network training. This paper presents a distributed model-parallel training framework that enables training large neural networks on small CPU clusters with low Internet bandwidth. We build upon the adaptive sparse training framework introduced by the SLIDE algorithm. By carefully deploying sparsity over distributed nodes, we demonstrate several orders of magnitude faster model parallel training than Horovod, the main engine behind most commercial software. We show that with reduced communication, due to sparsity, we can train close to a billion parameter model on simple 4-16 core CPU nodes connected by basic low bandwidth interconnect. Moreover, the training time is at par with some of the best hardware accelerators.
\end{abstract}

\section{Introduction}

A recent report~\cite{chapel_2020} highlights that 76\% of cloud computing resources that are paid for sit idle and are never used. This wasted computing exceeds 17.6 billion USD in 2020, or about 48 million USD everyday. In contrast, the cost of training GPT-3 from scratch is estimated to be about 12 million USD. This paper aims at designing efficient machine learning training algorithms and distributed implementations that can enable these wasted cloud computing resources to train (or refresh) heavyweight neural networks. It is beyond doubt that if we can extract AI from these wasted idle compute cycles, it could change the economics of Deep Learning in industry by dramatically reducing the cost associated with building models.    

However, to extract value from these idle cloud computing resources, our existing neural network training algorithms and state-of-the-art implementations must be transformed. Next, we describe some of the unique settings and constraints that neural network training algorithms must meet to utilize these idle compute cycles. 

\noindent \textbf{Leveraging a Large Number of Few-core CPU Nodes:} A significant fraction of the cloud consists of general-purpose few-core (4-16) CPUs with very limited local parallelism. However, there are a large number of computing nodes available. Distributed computing is therefore imperative to fully capitalize on the massive compute power available in a typical cloud.  

\noindent \textbf{Model Parallelism:}  Most cloud clusters operate on Virtual Machines (VMs) that facilitates automatic fault tolerance and share resources including data and files. However, the memory capacity of these VMs is not high. We can see typical computing nodes with about 8-16 GB of available memory, and the available memory is further reduced due to several legacy VM software running on them. As a result, we cannot expect a large neural networks model with parameters, auxiliary variables, and outputs to fit in the CPU node. Furthermore, due to limited parallelism on each node, our best hope to scale neural network training is to leverage parallelism in distributed settings. To enable considerable model training, we need to shard the parameters across distributed nodes. These requirements also prohibit the use of Federated Learning~\cite{fedLearning} techniques, which is meant for tiny models that can fit on any compute nodes. 

\noindent \textbf{Low Bandwidth Communication between Computing Nodes:}   One of the most limiting constraints of a typical compute cluster is low communication bandwidth between nodes. This constraint is most challenging for modern neural network training, which requires frequent and heavy communication between all the computing nodes. We see typical communication bandwidth in the range of $1$ to $100 \ Gbps$. Unfortunately, even the best distributed training algorithms do not cater to these constraints, and the associated implementations are prohibitive in the settings mentioned above.



\textbf{Current State of Model Parallelism for Training Large Neural Networks}: Model parallel training of neural networks is a very communication-heavy operation. Most of the current frameworks shard the layers of the neural network. As a result, every feed forward and backpropagation update requires fast communication equal to the size of the network per sample of the data. The communication cost multiplies by the batch size. No wonder, to make such expensive communication practical, the community is pushing the boundaries of hardware and investing in ultra-high bandwidth data transfer technologies such as NVLink~\cite{nvlink}. Unfortunately, such expensive communication is infeasible for the problem we are trying to address, and there is no other alternative available. To the best of our knowledge, this paper provides the first such alternative algorithm and implementation. 

We cannot afford the communication cost of existing training algorithms, and we need a significant departure from dense matrix multiplication-based backpropagation algorithm to reduce the computations. It should be noted that we are looking at model parallel training. Popular ideas of compression to minimize communication cost applies to data parallel training, and are not applicable in the current setting.  

\textbf{Sparsity and the SLIDE Algorithm:} There is a very appealing line of work~\cite{AdaDropout, AdaDOBa, spring2016scalable} that uses hash table to perform extreme adaptive dropouts. The resulting implementation, named as SLIDE~\cite{slide}, demonstrated extremely sparse computations ( $> 95\%$ adaptive sparsity) as well as gradient update associated with every data sample. When deployed on multi-core CPUs, the drastic reduction in computations was capable of outperforming standard backpropagation even when accelerated with top-notch GPUs. The downside was that the sparsity was dynamically chosen by the hash table and, hence, dependent on the input. Thus, utilizing memory coalescing~\cite{daghaghi2021accelerating} was a challenge.

The most appealing aspect of the SLIDE algorithm was that the computation and memory access associated with processing any input and associated gradient update was 95\% sparse. Thus, there is hope for a significantly reduced communication footprint if we can obtain a distributed implementation. 

\textbf{Our Contributions:}  We present a distributed model parallel training framework to train large neural networks on small CPU clusters with low Internet bandwidth. We capitalize on sparsity of the SLIDE algorithm to reduce the communication footprint associated with model parallel training of neural networks by orders of magnitude. Our contribution is both novel algorithm and system design. We provide an MPI (Message Passing Interface) \cite{mpi40} based implementation that makes novel choices of determining how to shard neural network and distribute hash table computation over multiple nodes to ensure load balancing and reduced communication. We also provide a rigorous evaluation of D-SLIDE, where we demonstrate several orders of magnitude faster model parallel training than Horovod~\cite{horovod}, the main engine behind commercial software that supports model parallelism, including Amazon Sagemaker \cite{sagemaker}, Azure Databricks \cite{AzureDatabricks}, and Databricks Spark \cite{horovod.spark}. We show that with reduced communication, due to sparsity, we can train a near-billion parameter model on simple 4-16 core CPU nodes connected by basic low bandwidth interconnect. Moreover, even with a severely restricted platform, the training time is at par with some of the best hardware accelerators. We are very excited about the positive results and their capability to change the whole economics of neural network training.  


\section{Background}
\subsection{The SLIDE Algorithm}

Sparsity plays a key role in many modern deep learning workloads. With the recent advance in using Locality Sensitive Hashing (LSH) \cite{LSH, SRP} as a sampler~\cite{spring2017new, spring2021mutual}, SLIDE \cite{slide} provides a solution to train neural networks on CPUs efficiently. SLIDE leverages LSH to adaptively select very few neurons having large activations. The SLIDE algorithm works in three phases. Phase 1 is the \textbf{initialization} phase. In addition to initializing weights of the model, SLIDE indexes weight vectors in each layer to corresponding key-value based hash tables for efficient retrieval. This creation and indexing are one-time operations. During training, the network iteratively goes to feed forward and backpropagation stages. In the \textbf{feed forward} stage, we query each layer's hash tables with the input to the layer to retrieve few neurons relevant to the input. We then compute the activations of the retrieved neurons only. Activations of other neurons are treated as 0 and are no operations (NoOps). The sparse activation vector for every layer now becomes the transformed input that is fed to the next layer. In the \textbf{backpropagation} stage, errors and gradients are backpropagated to only the active neurons. For more details, please refer to \cite{slide}. Later, \cite{daghaghi2021accelerating} further improves the performance of the SLIDE algorithm by leveraging vectorization and memory coalescing techniques. 

\begin{algorithm}

\begin{algorithmic}
\STATE \textbf{Input:} Batch size $B$, Epochs $e$, Layers $n$ \\
\STATE Initialize layers with randomly generated weights. \\
\FOR{$i \gets 1$ to $n$}
    \STATE Initialize hash tables
    \STATE Hash weights to hash tables
\ENDFOR
\FOR{$i \gets 1$ to $e$}
    \FOR{$j \gets 1$ to $B$}
        \FOR{$k \gets 1$ to $n$}
            \STATE Retrieve Active Neurons $AN_k$
            \STATE $Act_{k} \gets $ Feed Forward($Act_{k-1}, AN_k$)
            \STATE \textbf{Synchronize activations across nodes}
        \ENDFOR
        \STATE Compute Errors $Err_n$
        \STATE \textbf{Synchronize errors across nodes}
        \FOR{$k \gets n \ \KwTo \ \ 1$}
            \STATE $Err_{n-1} \gets $ Backpropagation ($Err_n$)
            \STATE \textbf{Synchronize gradients across nodes}
        \ENDFOR
    \ENDFOR
    \STATE Update weights $w_1, w_2, \dots, w_n$ with Adam
    \STATE Rebuild hash tables and hash functions  
\ENDFOR
 \end{algorithmic}
  \caption{D-SLIDE Training}
  \label{alg:train}
\end{algorithm}

\subsection{Communication Costs of Model Parallel Training }

A significant bottleneck in distributed deep learning is communication, particularly where the network's bandwidth limits us. To understand this bottleneck, consider a layer of $L$ neurons, batchsize $B$, distributed over $n$ computing nodes. In model parallel training, existing packages (such as Horovod \cite{horovod}) would divide up the layer into $n$ partitions, and each node would have $\frac{L}{n}$ neurons. In the current paradigm, activations for all $\frac{L}{n}$ neurons would need to be communicated to the other nodes, and we need to repeat this for every data point in the batch. We need to share $B \times L$ activations for each batch. Similarly, we would need to communicate $B \times L$ errors and partial gradients during the backpropagation phase. This $O(L)$ communication per data sample per update is simply prohibitive for large models in our setting where we have low bandwidth interconnect. We will introduce our solution D-SLIDE in the following section to address this bottleneck.


\section{Our Proposal: D-SLIDE}

We propose D-SLIDE, a distributed version of hash table-based SLIDE algorithm, to address the communication bottleneck discussed in the previous section. We consider the same setting where we have a fully connected layer of $L$ neurons, batchsize $B$, and $n$ nodes. Due to the highly sparse nature of the SLIDE algorithm, it only retrieves $K$ active neurons, where $K \ll L$ in a layer. A sparse activation means that we would only need to communicate $B \times K$ activations during the feed forward stage, as the others are NoOps, and similarly $B \times K$ gradients during backpropagation. Since $K$ is orders of magnitude smaller than  $L$ in practice, we can reduce the communication cost significantly by leveraging the sparsity that the SLIDE algorithm offers. Algorithm \ref{alg:train} delineates the training procedure.

\begin{figure*}
\centering
\includegraphics[width=\linewidth]{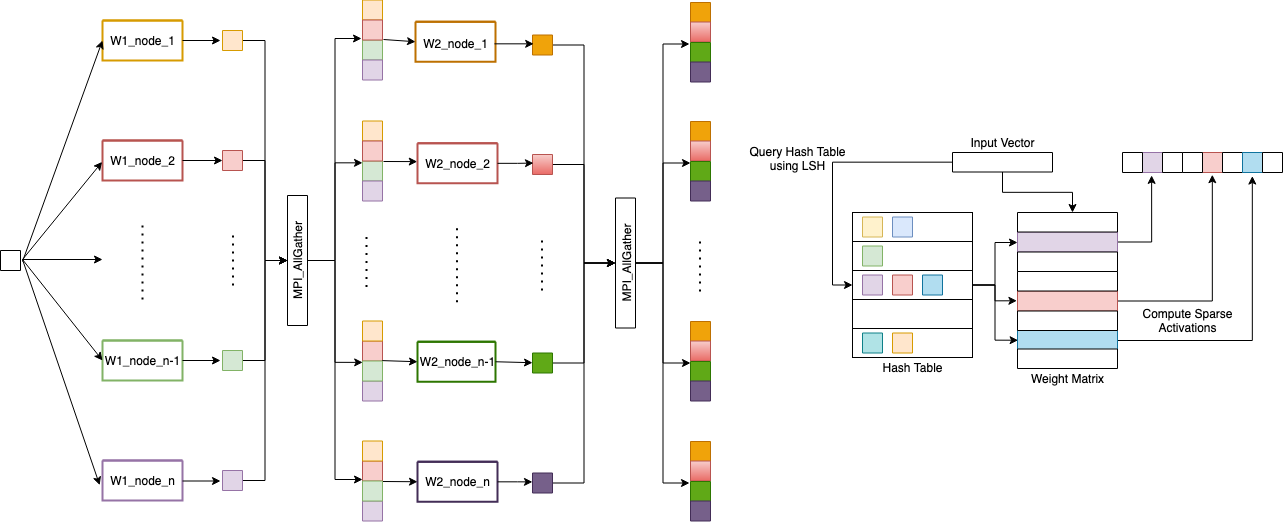}  
\caption{On the left side, the figure shows our collective communication design for gathering activations during forward propagation. After processing each batch of data, we communicate activations or gradients with all other nodes and obtain a copy of full activations or gradients for all neurons in a layer. On the right side, we illustrate the process of selecting active neurons using LSH and computing sparse activations. Note that in practice, we use multiple hash functions and hash tables and compute activations for the union of neurons retrieved from each hash table. We simplify the process to only use one hash table in the diagram for better visual clarity.}
\label{fig:diagram_comm}
\end{figure*}

\subsection{Sharding the Model and Hash Tables}

The architecture of D-SLIDE is illustrated in Figure~\ref{fig:diagram_comm}. We shard the network similar to the standard way. The layers consisting of neurons are divided over the computing nodes. Each computing machine owning the neurons has all its weights. The activation of the neuron is completely calculated by the owner of the neuron. We only communicate input and activations across the machines, and the weight vectors are never communicated across nodes during training. The network distribution is straightforward. However, the network selects active neurons by querying the input to a set of hash tables. 

We also shard the hash tables so that every node owns the hash tables needed to query its subset of neurons. 
In particular, each computing machine contains its own set of hash tables that was indexed by the partition of neurons that the machine owns. The hash tables are constructed with weight vectors and locally updated with a frequency schedule. Each computing unit hashes the input to sample active neurons from local hash tables during training. Since the neurons are partitioned disjointedly across machines, our design choice allows us not to incur any additional communication cost. 

\subsection{Use of Parallelism}
In the SLIDE algorithm, HOGWILD-style batch parallelism \cite{niu2011hogwild} is applied during training at the network level where each data sample in a batch performs feed forward and backpropagation in parallel without any synchronization in between. In D-SLIDE, since the layers need to be distributed, we have to synchronize activations and gradients after each layer. We then apply HOGWILD-style parallelism at the layer level and set a barrier after each layer via synchronized MPI Allreduce and Allgather operations.

\subsection{Load Balancing the Computations}
A key challenge in deploying the SLIDE algorithm in a distributed setting is load balancing. In the SLIDE algorithm, the neuron selection process does not control how many neurons get selected from each machine. Even though we only need to communicate only $K \ll L$ communications per data sample, if all those $K$ neurons come from one machine, we end up having most computations only on that machine, which will hurt the performance.

To ensure load balancing, we choose to force each machine to select $K/n$ number of active neurons. This forcing is a necessary change we made towards the original SLIDE algorithm. However, we found that this does not affect the performance in practice. During the initialization phase, we randomly initialize the weight vectors. Since $L \gg n$, each node should have similar number of neurons with high activations. Therefore, the algorithm's performance would not be affected if we start forcing the nodes to select the same number of neurons from the beginning. In the experiment section, we show a rigorous evaluation demonstrating that our implementation does not hurt the model's accuracy.

\subsection{Enabling Large Batch Size Training}
One of the bottlenecks of GPUs is their memory bandwidth. Even on the latest GPUs, training with a large batch size often leads to an out-of-memory error. This is because GPU, for utilizing parallelism, replicates a significant chunk of activations and errors for every element in the batch. On the other hand, CPUs have much larger memory at a much cheaper cost. Furthermore, our implementation only activates very sparse computations. The need for replication is orders of magnitude lower due to sparsity. As a result, our method can train large models with large batch sizes. \cite{you2017large}, \cite{you2019large}, \cite{you2019large2}, and \cite{you2017scaling} have shown that training with large batch size may lead to better accuracy. However, previous lines of work in exploring large batch size training were mostly performed on large GPU-clusters not accessible to most of the users. The experiment section shows that we can train a large network with a large batch size, a computation that otherwise cannot fit into a GPU.

 


\subsection {Other Implementation Details}
Distributed SLIDE is designed to be a framework for efficient deep neural network training on CPUs. We made the interface easily extendable to various kinds of neural networks, including but not limited to DLRM \cite{naumov2019deep} and GCN \cite{defferrard2017convolutional}. We shard our models evenly across nodes. Each node contains a partition of the neurons for each layer. During the forward propagation phase, active neurons are selected independently on each node in parallel, and parts of the activation array are computed independently before being merged.
Similarly, parts of the gradient array are computed in parallel on each node before merging during the backpropagation phase. Each active neuron contains a weight array and a bias array of the size of the previous layer. When the model is trained with a momentum-based optimizer, such as Adam, the model also maintains a momentum array and a velocity array for each corresponding weight and bias array. 

\subsection{Interface for Layer}
To build a framework that can be extended to different layers, we abstracted three operations essential to our workflow: feed forward, backpropagation, and active neurons selection. The feed forward function is further divided into two types. The first type of feed forward function accepts processed sparse datasets, which contain arrays of indices, values, and labels. Activations are computed and stored in a snapshot of the layer that includes a two-dimension array of active neurons and their corresponding activations and backpropagated errors for each data sample in a batch. The second type of feed forward function accepts the previous layer's snapshot as input and updates its snapshot to feed into the next layer. We decouple the neuron selection process from the forward propagation process. This enables us to support dense feed forward layers without additional code by simply checking if the layer sparsity is 1.


\subsection{Early Stopping for Fast Neuron Sampling}
Sparsity is a parameter of our algorithm that determines how many neurons to sample. If the number of neurons sampled from the hash tables exceeds each node's sampling budget, we perform reservoir sampling \cite{reservoir} to control the number of active neurons selected. When the number of sampled neurons is less than each node's sampling budget, we provided users with two options to either fill the rest of the sampling budget with uniformly sampled neurons or stop with the current set of neurons. 


\subsection{MPI for Peer to Peer Sparse Communication and Synchronization} \label{EffComm}
In model parallel training, we synchronize the activations after processing a batch of data through a layer in the feed forward stage and synchronize the gradients after backpropagating a batch through a layer. To leverage the collective communication mechanisms in MPI (E.g. Allgather(v), Allreduce(v)), we create a snapshot for each layer in our network, where we store information that needs to be synchronized, including active neurons, active neuron counts, activations, and gradients. This design choices allow us to decouple communication from concrete feed forward and backpropagation operations, making it flexible to support different types of layers in the future. In our design, to take full advantage of HOGWILD-style batch parallelism, we use one rank per node and cannot parallelize our communication as MPI cannot differentiate the threads within a node. MPI calls act as critical sections in our algorithms. Supporting parallel MPI calls in our scenario is of independent research interest and would provide further speedup, especially when the algorithm is deployed to many nodes.

\section{Experiment}
This section will provide an in-depth comparison between our method and the state-of-the-art distributed training methods on GPUs. To fully unleash the power of our approach, we focus on multilayer perceptrons (MLP) on the scale of hundreds of millions and billions of parameters to demonstrate the potential of D-SLIDE. We perform extensive tests of our model on large publicly available extreme classification and natural language processing (NLP) datasets. 

We show that our method exhibits desirable scalability when deployed on a large number cheap computing nodes and can quickly train billion parameter network that is difficult to fit into GPU memory. We also benchmark our method against TensorFlow on the latest models of Nvidia GPUS, V100, and A100. The aim is to get a sense of where the algorithm on cheap computing resources stands compared to best practices. 

We further examine the scalability of our method by deploying our code in settings with both high and low bandwidth interconnect. We compare our approach against the state-of-the-art distributed training framework Horovod \cite{horovod} and show that we achieve orders of magnitude faster training time due to our reduced communication.

\subsection{Models and Hyper-parameters} \label{hyperparam}
We perform our experiments on two-layer fully connected networks where the input and output dimension can be found in table \ref{tab:datasets} and the hidden layer size is specified in each experiment.

In addition to the standard parameters in deep learning, our hashing-based method introduces several new hyper-parameters. We use $K$ to denote the number of hash functions and $L$ to denote the number of hash tables. $K$ is set to 6 for experiments performed with Densified Winner Take All Hashing (DWTA) \cite{dwta} and 9 for experiments performed with Sparse Random Projections (SRP) \cite{SRP}. $\beta_1$ and $\beta_2$ in Adam optimizer are set to $0.9$ and $0.999$, $\epsilon$ is set to $10^{-8}$, and learning rate is set to be $0.0001$. Note that these parameters are chosen after extensive testing of both our method and baseline. We vary the hidden layer size and batch size to understand our method's performance in various settings fully. The details of hidden layer size and batch size are discussed in the following experiments, respectively.\\

\begin{table*}[t]
\centering
\caption{The table shows the statistics of the datasets. Here the feature sparsity represents the average percentage of non zero entries in the feature vector.}
\begin{tabular}{|c|c|c|c|c|c|}
\hline
& Feature Dim & Feature Sparsity & Label Dim & Train Size & Test Size \\ \hline
Amazon670K & 135909 & 0.055\% & 670091 & 490449 & 153025 \\ \hline
Text8 & 253855 & 0.0004\% & 253855 & 13604165 & 3401042\\ \hline
 
\end{tabular}

\label{tab:datasets}
\end{table*} 

\subsection{Datasets}
We perform our experiments on publicly available datasets Amazon670K and Text8. 

Amazon670K is an extreme classification dataset released in a Kaggle competition. We use the public version provided in \cite{Bhatia16}. Each data sample is a sparse vector representation of a product. The labels are related products a user might also purchase. Each data sample is of dimension 135909, and there are 670091 distinct labels in the dataset.

Text8 \cite{text8} is an NLP dataset extracted and processed from 100 million tokens in English Wikipedia. The dataset contains a vocabulary of 253K words. We process the dataset as described in \cite{daghaghi2021accelerating}. 

Refer to table \ref{tab:datasets} for detailed statistics of the datasets.

\subsection{Infrastructure} \label{high_infra}
We perform our experiments on multi-node blade clusters with Intel(R) Xeon(R) Gold 6230 CPUs. Each CPU has a clock rate of 2.10GHz. Each node has 80 cores and 160 threads. To demonstrate our performance on a more moderate server, we restrict to use only 40 cores (80 threads) and 2 nodes for all experiments (Except for the scalability experiments in table \ref{tab:multi-node-scalability} and \ref{tab:low-bandwidth-scalability} where we explicitly state the number of cores/threads used). Our nodes are connected via Intel OmniPath, and it has a bandwidth of 100Gb/s.

Nvidia Tesla V100 and A100 are used for baseline experiments without interference from other processes. V100 has 32GB of memory on each GPU, and A100 has 40GB. Except for experiments reported in table \ref{tab:measures} and figure \ref{fig:Bar}, where multiple GPUs are required to fit the model, we perform our experiments on a single GPU.

\begin{figure*}
\centering
\begin{subfigure}{0.38\linewidth}
  \centering
  \includegraphics[width=\linewidth]{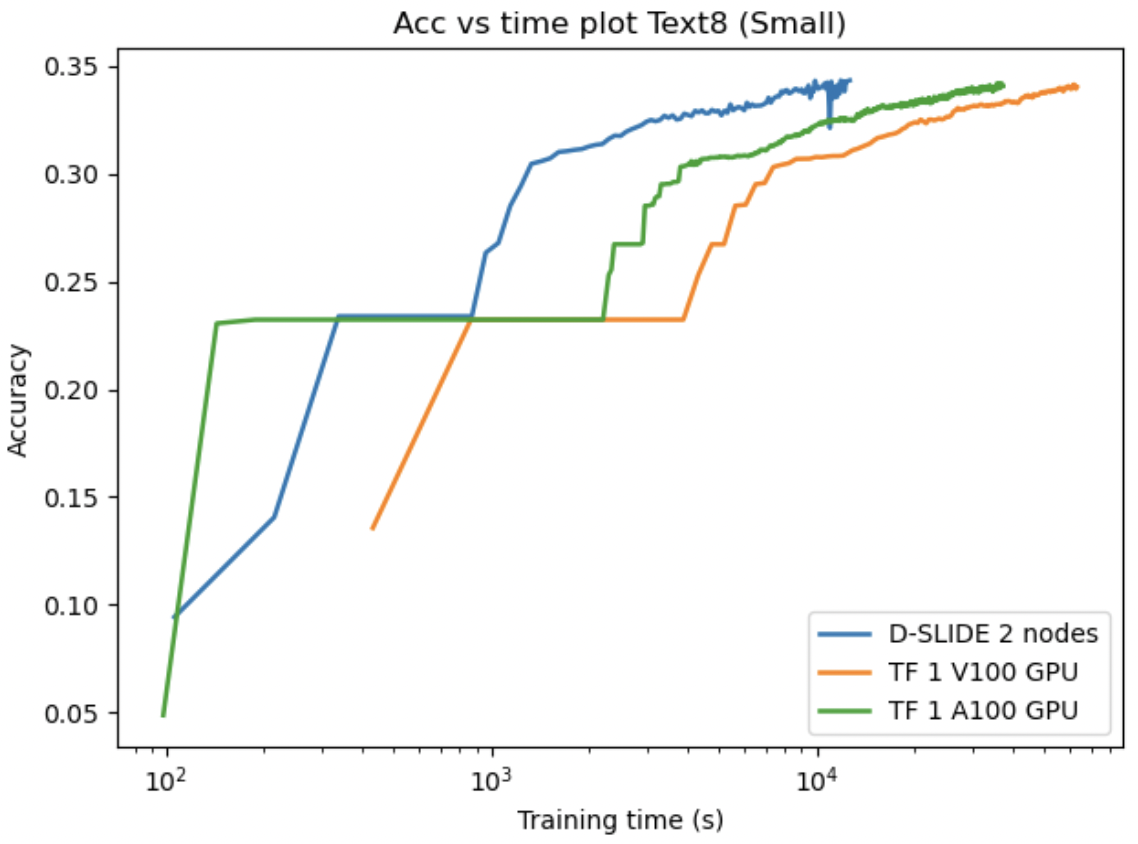}  
  \label{fig:Text8Conv}
\end{subfigure}
\begin{subfigure}{0.38\linewidth}
  \centering
  \includegraphics[width=\linewidth]{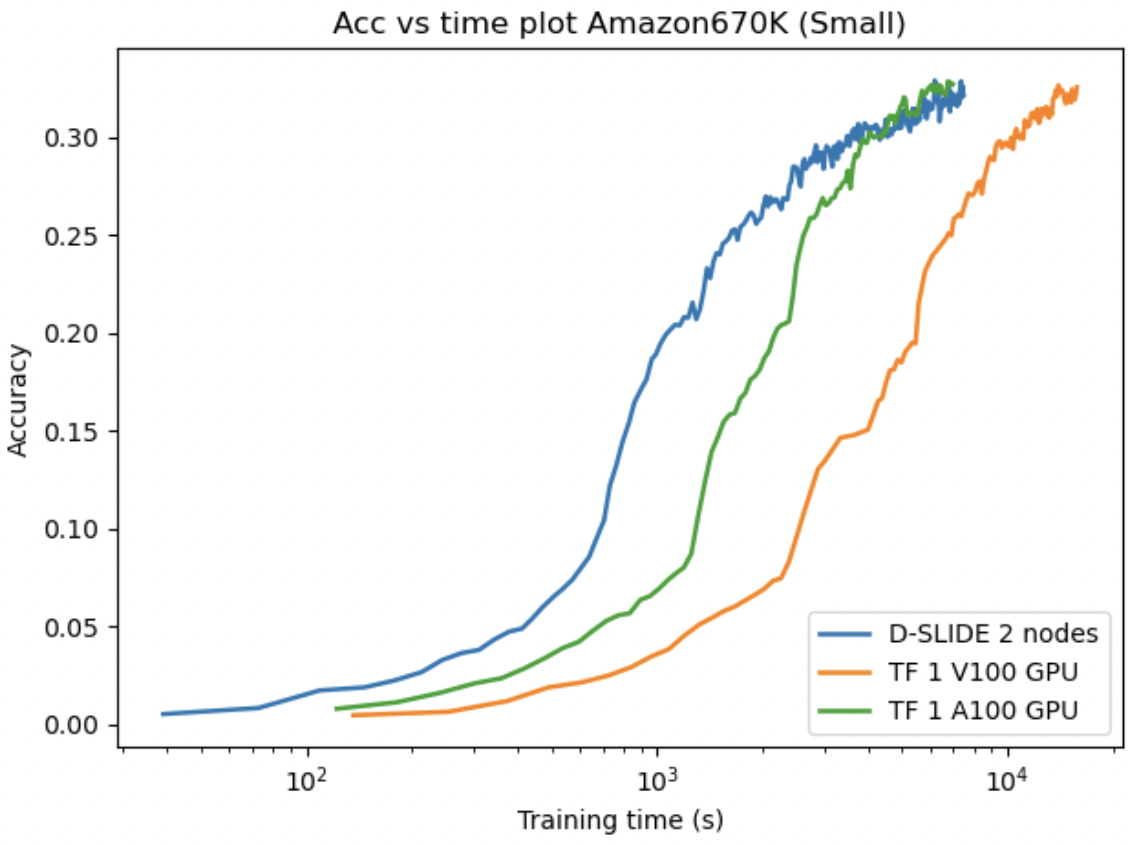}  
  \label{fig:Amazon670KConv}
\end{subfigure}
\caption{In this figure, we show the convergence plot of D-SLIDE distributed over \textbf{2 nodes} against Tensorflow on \textbf{1 V100 GPU} and \textbf{1 A100} GPU without any distribution, respectively. The model is small enough to fit on a single GPU. On the left side, we show the plot for text8 dataset; on the right side, we show the plot for Amazon670K dataset. D-SLIDE distributed over two nodes converges \textbf{4.0x} faster than TensorFlow on an A100 GPU and \textbf{7.0x} faster than a V100 GPU on text8 dataset and converges as fast as an A100 GPU and \textbf{2.7x} faster than a V100 GPU on Amazon670K dataset. We perform this experiment on a cluster with 100 Gbps interconnect bandwidth.}
\label{fig:Convergence}
\end{figure*}

\subsection{Small Model Experiment: Comparison with Tensorflow with Single GPU Acceleration}
In this section, we provide a head-to-head comparison with Tensorflow \cite{tensorflow} accelerated with powerful GPUs on small models that fit those GPUs without the need for any distribution. The aim here is to ballpark D-SLIDE training time with the best possible memory on GPUs even on a much smaller scale. 

Figure \ref{fig:Convergence} shows the convergence plot against Nvidia V100 and A100 GPUs on Amazon670K and Text8 dataset. On Text8 dataset, D-SLIDE on 2 nodes converges \textbf{4.0x} faster than A100 GPU and \textbf{7.0x} faster than V100 GPU. On Amazon670K dataset, our method converges as fast as A100 GPU and \textbf{2.7x} faster than V100 GPU.



\begin{table}[t]
\caption{This table shows the scalability of D-SLIDE. It provides the relative speedup of our method with varying parallelism in the form of nodes and cores. We measure the per epoch training time and limit the number of available threads and nodes. We use the time obtained when running on 1 node and 4 threads as the baseline and report the comparative speedup across various number of threads and nodes. We perform this experiment on a cluster with 100 Gbps interconnect bandwidth.}
\centering
\begin{tabular}{|c|c|c|c|c|c|}
\hline
 & \multicolumn{4}{|c|}{ Number of threads on each node} \\ \hline
\# of nodes & 4 & 8 & 16 & 32 \\ \hline
1 & 1x & 1.81x & 2.26x & 3.29x\\ \hline
2 & 1.31x & 2.08x & 3.07x & 4.12x\\ \hline
4 & 2.12x & 3.11x & 3.98x& 4.55x\\ \hline
 
\end{tabular}

\label{tab:multi-node-scalability}
\end{table} 

\subsection{Scalability Experiment: D-SLIDE with more Cores and Nodes}
In this experiment, we show that our method provides desirable scaling capabilities. We conducted our investigation on Amazon-670K dataset with batch size and hidden layer size set to be 1024. Table \ref{tab:multi-node-scalability} shows the speedup on a various number of nodes and cores. We demonstrate that when the number of nodes is fixed, our method has strong scaling capabilities as we increase the number of cores on each node. When the number of cores on each node is fixed, we can achieve significant speedup by increasing the number of nodes. 

Table \ref{tab:multi-node-scalability} shows when the number of threads on each node is small, D-SLIDE exhibits better scalability on multiple nodes compared to when deploying on 32 threads. Under our batch parallel paradigm, doubling the number of threads theoretically doubles the number of data samples that can be processed simultaneously, enabling us to gain more speed. However, as discussed in \ref{EffComm}, the MPI collective communication operations cannot be parallelized. Therefore, when deployed on a large number of threads, we see a diminishing speedup on multiple nodes since the communication cost becomes more significant. However, this would not be an issue for most large-scale deep learning workloads. Modern CPU clusters with 4 or 8 nodes have terabytes of memory, enabling us to train models with tens or hundreds of billions parameters.

\begin{figure*}

\centering
\begin{subfigure}{0.35\textwidth}
  \centering
  \includegraphics[width=\linewidth]{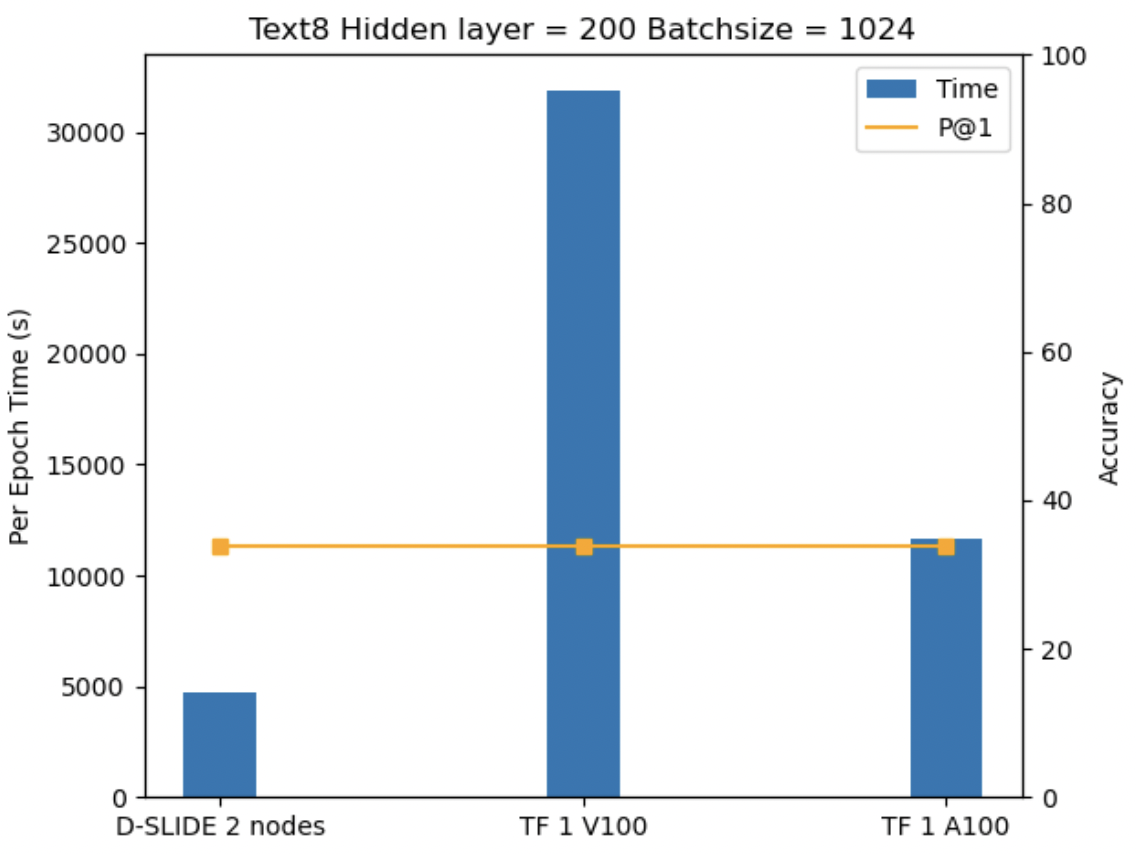}  
  \label{fig:text8bar}
\end{subfigure}
\begin{subfigure}{0.35\textwidth}
  \centering
  \includegraphics[width=\linewidth]{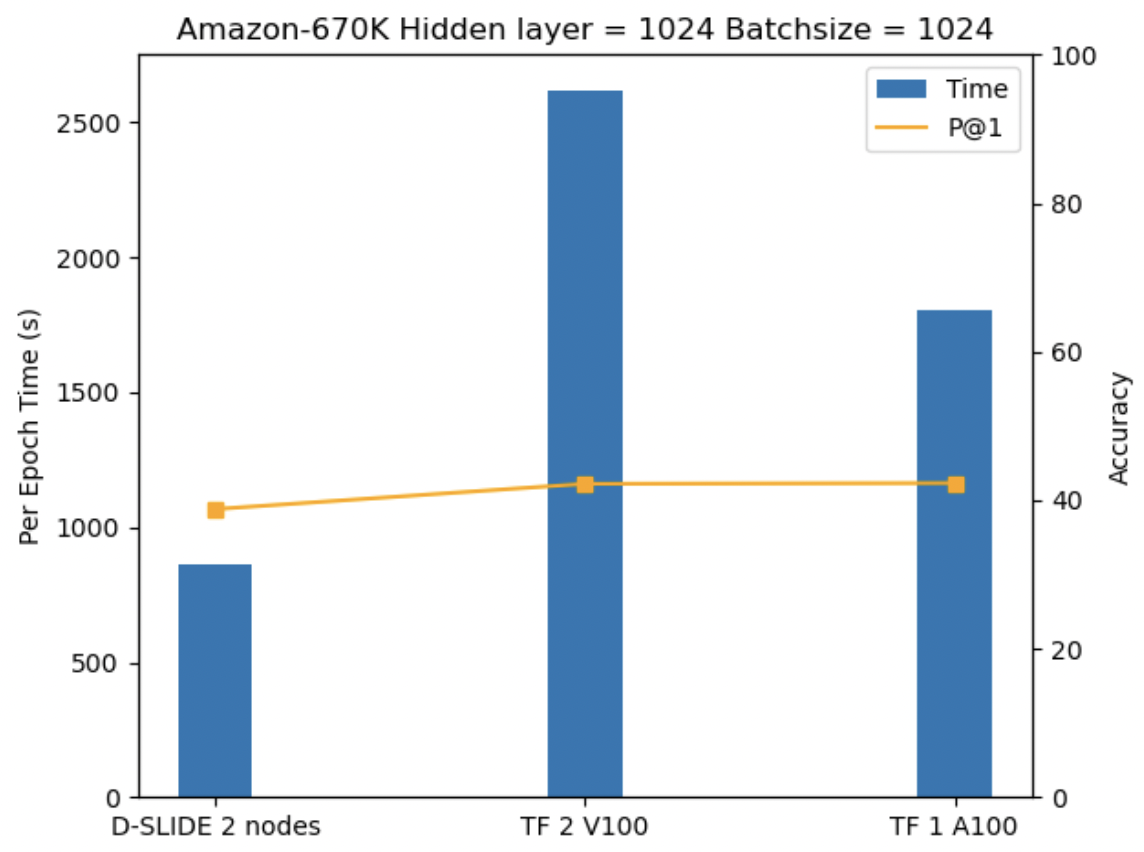}  
  \label{fig:amzbar}
\end{subfigure}
\caption{On the left side, the figure shows that D-SLIDE on 2 nodes is \textbf{2.5x} faster than 1 Nvidia A100 GPU and \textbf{6.7x} faster than 1 V100 GPU in terms of per epoch training time when training on the large batch size on Text8 dataset and converges to the same accuracy. On the right side, the figure indicates that D-SLIDE on 2 nodes is \textbf{2.1x} faster than Nvidia 1 A100 GPU and \textbf{3.0x} faster than 2 V100 GPUs per epoch on the large batch size on Amazon670K dataset and achieves accuracy close to TensorFlow. We perform this experiment on a cluster with 100 Gbps interconnect bandwidth.}
\label{fig:Bar}
\end{figure*}

\begin{table*}
\caption{The table shows the time required to reach a certain target accuracy when training on large batch size on Amazon670K and Text8 dataset. We notice that on Amazon670K dataset our method even when using two distributed CPUs is only slightly slower than TensorFlow on A100 and has comparable or better performance than on V100. On Text8 dataset, our method outperformed both A100 and V100 GPUs by a large margin. Note that though A100 is 2.5x faster than V100 in terms of per epoch time, the time A100 takes to reach the target accuracy is similar or even more than the time V100 takes on Text8 dataset. Since we are running the same baseline code on both hardware, we speculate that A100's internal mixed precision optimization hurts its convergence speed. We perform this experiment on a cluster with 100 Gbps interconnect bandwidth.}
\centering
\resizebox{\columnwidth}{!}{%
\begin{tabular}{|c|c|c|c|c|c|c|c|c|}
\hline
\multicolumn{4}{|c|}{Method} & \multicolumn{2}{|c|}{TF A100} & \multicolumn{2}{|c|}{TF V100} & D-SLIDE 2 nodes \\ \hline
Dataset & Hidden size & Batchsize & Target Acc & Time & \# GPUs  & Time & \# GPUs & Time\\ \hline
Text8 & 200 & 1024 & 33\% & 45182s & 1 & 46928s & 1 & 6290s\\ \hline
Text8 & 200 & 2048 & 33\% & 110081s & 1 & 73290s & 1 & 9374s\\ \hline
Amazon670K & 1024 & 1024 & 35\% & 3238s & 1 & 4278s & 2 & 4382s\\ \hline
Amazon670K & 1024 & 2048 & 35\% & 3670s & 2 & 5952s & 5 & 4275s\\ \hline
 
\end{tabular}
}
\label{tab:measures}
\end{table*} 

\subsection{Large Model Experiment: Comparison with Distributed GPU Baselines}
The following experiments investigate training a large neural network that requires model parallelism with GPUs because of model size or batch size. Table \ref{tab:measures} shows the hidden layer size, and batch size for each experiment and how many GPUs are required to fit a model of such size. As in the previous experiment, D-SLIDE is deployed on 2 nodes. Again, the aim here is to understand the trade-offs of using D-SLIDE on CPU for model parallel training.



Table \ref{tab:measures} shows that on Amazon670K dataset, our method is comparable or faster than 2 V100 GPUs to reach target accuracy on large batch sizes and is slightly slower than A100, and on the Text8 dataset, our method converges to target accuracy \textbf{7.5x - 7.9x} faster than V100 GPUs and \textbf{7.2x - 11.7x} faster than A100 GPUs. Figure \ref{fig:Bar} shows that our method has a significantly faster per epoch training time compared to both A100 and V100 GPUs and achieves convergence accuracy close to GPUs on Amazon670K and comparable accuracy on the Text8 dataset. 

Figure \ref{fig:Bar} shows the bar chart where the left y-axis and right y-axis represent per epoch training time and test accuracy, respectively. On the left side, the figure indicates that D-SLIDE on 2 nodes is \textbf{2.5x} faster than 1 Nvidia A100 GPU and \textbf{6.7x} faster than 1 V100 GPU in terms of per epoch training time when training on the large batch size on Text8 dataset and converges to the same accuracy. On the right side, the figure shows that D-SLIDE on 2 nodes is \textbf{2.1x} faster than 1 Nvidia A100 GPU and \textbf{3.0x} faster than 2 V100 GPUs per epoch on the large batch size on Amazon670K dataset and achieves accuracy close to TensorFlow.

 


\begin{table*}[t]
\caption{This table shows the scalability of our method with few core CPUs and a low bandwidth communication. We measure the per epoch training time and compare it against the widely-adopted Horovod package, which is the de facto state-of-the-art that is used by a variety of distributed training framework, including but not limited to Amazon Sagemaker and Apache Spark.}
\centering
\resizebox{\columnwidth}{!}{%
\begin{tabular}{|c|c|c|c|c|c|c|c|}
\hline
& & & \multicolumn{4}{|c|}{ Number of nodes} \\ \hline
Method & Cores & Internet Bandwidth (Gbps) &1 & 2 & 4 & 8 \\ \hline
D-SLIDE & 4 & 1 & 16781s & 9241s & 6435s & 5417s\\ \hline
D-SLIDE & 16 & 1 & 6706s & 4166s & 3260s & 3840s\\ \hline
Horovod & 16 & 1 & 7388s & $2.75 \times 10^5$s & $2.60 \times 10^5$s & $2.82 \times 10^5$s \\ \hline
Horovod & 16 & 100 & 6628s & 37417s & 40222s & 41102s \\ \hline
 
\end{tabular}
}
\label{tab:low-bandwidth-scalability}
\end{table*} 

\subsection{Our Goal Task: Training 800M Parameter Model on Few Core CPUs with Low Communication Bandwidth and Comparison with Horovod}
This section demonstrates that our approach can train large models in a distributed setting even when the Internet bandwidth among the nodes is slow ( $\le$ 1Gbps). This bandwidth is orders of magnitude slower than the connection deployed on most distributed clusters used for deep learning. 

We compare our method against Horovod, which is de facto state-of-the-art distributed training framework available. Horovod is the backend for most of the distributed neural network training platforms, including but not limited to Amazon Sagemaker \cite{sagemaker}, Databricks Spark \cite{horovod.spark}, and is well integrated to Pytorch \cite{pytorch}, Tensorflow \cite{tensorflow}, and Keras \cite{keras}. Since we also use MPI for our communication among nodes, we deploy Horovod with MPI for fair comparison. 

\noindent\textbf{Parameters}: In this experiment, we consider the setting where our cluster has low interconnect bandwidth (1Gbps), low number of cores (4-16 on each node), and low memory. Therefore, compared to our previous scalability experiment, we chose a smaller hidden layer size 256 and set the batch size to 2048. Our method hits peak memory usage of 18GB when deployed on 1 node and gradually decreases as we deploy our approach on more nodes. The rest of the hyper-parameters are identical to the setting in section \ref{hyperparam}.

\noindent\textbf{Infrastructure}: In this experiment, we deploy our code on clusters with Intel(R) Xeon(R) CPU E5-2680 v3 @ 2.50GHz processors. The Interconnect between nodes has a bandwidth of 1Gbps. The last line in table \ref{tab:low-bandwidth-scalability} is run on the server mentioned in section \ref{high_infra}. 

\noindent\textbf{Results}: Table \ref{tab:low-bandwidth-scalability} shows the comparisons between our method and Horovod in low bandwidth settings. The table clearly demonstrates that deploying any reasonably sized model to multiple nodes with low internet bandwidth is prohibitive due to communication bottleneck. In contrast, with $99\%$ communication compression, our method still shows reasonable speedup when deployed on multiple nodes, especially when the computation resource on each node is frugal. With 16 cores and 1 Gbps bandwidth, our method is \textbf{66 - 80x} faster than Horovod when deployed on multiple nodes and is even an order of magnitude faster than Horovod when Horovod is deployed on a cluster with high bandwidth.


\section{Conclusion}
This paper presents D-SLIDE, a distributed model-parallel training framework to train large neural networks on small CPU clusters with low Internet bandwidth. We provide rigorous experimental results demonstrating several orders of magnitude faster model parallel training when deployed in low-resource settings. We show that with reduced communication, we can train near-billion parameter models on superficial 4-16 core CPU nodes with low bandwidth interconnect by leveraging sparsity in the SLIDE algorithm. We provide an MPI-based implementation that shards neural networks and distributes hash tables over multiple nodes to ensure load balancing and reduced communication. We further show that the training time is at par with some of the best hardware accelerators, even with a severely restricted platform.

\bibliographystyle{unsrt}
\bibliography{main}

%





\end{document}